\documentclass[a4paper,12pt]{article} 
\usepackage[a4paper,top=7mm,bottom=7mm,left=7mm,right=7mm,footskip=0pt,headheight=0pt]{geometry}
\usepackage{graphicx}
\usepackage{epstopdf}
\usepackage{mathtools}
\usepackage{float}
\usepackage[font={color=black,small},labelfont={color=black,small,bf}]{caption}
\usepackage{subfig}
\usepackage{xfrac}
\usepackage{array}
\usepackage{natbib}
\usepackage{enumitem}
\usepackage{enumitem}
\usepackage{etaremune}
\usepackage{titlesec}
\usepackage[dvipsnames,table]{xcolor}
\usepackage{multirow}
\usepackage{ragged2e}

\titleformat{\section}{\bfseries\color{blue}}{\thesection.}{0.5em}{}[\titlerule]
\titlespacing{\section}{0pt}{10pt}{5pt}
\titleformat{\subsection}{\bfseries\color{black}}{\thesubsection.}{0.5em}{}
\titlespacing{\subsection}{0pt}{0pt}{5pt}

\author{
  A.~S.~G. Robotham$^{1}$
  \and
  Cullan Howlett$^{1}$\\
  \texttt{aaron.robotham@uwa.edu.au, cullan.howlett@uwa.edu.au}
  \\\\
  $^{1}$ICRAR, M468, University of Western Australia, Crawley, WA 6009, Australia\\
}

\title{A Short Research Note on Calculating Exact Distribution Functions and Random Sampling for the 3D NFW Profile}

\begin{document}

\maketitle

\pagestyle{empty}

\abstract{In this short note we publish the analytic quantile function for the Navarro, Frenk \& White (NFW) profile. All known published and coded methods for sampling from the 3D NFW PDF use either accept-reject, or numeric interpolation (sometimes via a lookup table) for projecting random Uniform samples through the quantile distribution function to produce samples of the radius. This is a common requirement in $N$-body initial condition (IC), halo occupation distribution (HOD), and semi-analytic modelling (SAM) work for correctly assigning particles or galaxies to positions given an assumed concentration for the NFW profile. Using this analytic description allows for much faster and cleaner code to solve a common numeric problem in modern astronomy. We release {\sc R} and {\sc Python} versions of simple code that achieves this sampling, which we note is trivial to reproduce in any modern programming language\footnote{{\sc R}: https://github.com/asgr/NFWdist; {\sc Python}: https://github.com/CullanHowlett/NFWdist}.}

\noindent
\section{Sampling the 3D NFW Profile}

It is a common situation in modern astronomy to require samples from the \citep[NFW;][]{NFW} profile. Typical applications include distributing particles within an $N$-body simulation, galaxies within halos in a halo occupation distribution (HOD) analysis, and assigning satellite galaxies that have lost tracking within a semi-analytic model (SAM). Efficiently distributing within a 3D NFW means interpreting the mass profile as a probability distribution function (PDF). All known applications the authors are aware of (either published, or in popular packages for handling halos) use one of two approaches:

\begin{enumerate}
\item Creating $x$ linear random Uniform samples between 0 and $R_{vir}$ and $y$ separate linear random Uniform samples between 0 and maximum differential mass as a function of the 3D radius (found at $R_s$, where the $M_{vir}$ normalised version of this distribution is the PDF form of the NFW mass profile). Samples below the expected differential mass for a given radius are kept, and the rest are rejected. This technique is commonly known as `accept-reject', and it is accurate but slow for creating random samples. Its efficiency is dependent on the fraction of the area sampled that sits below the PDF. If the area if mostly sparse then many more samples must be generated for the target number of samples to keep. In practice, for the NFW profile, the efficiency sits around the 50\%-70\% level. The {\sc Mock-Factory} package uses this technique\footnote{https://github.com/mockFactory/}\\
\item Creating the 3D mass integral as a function of radius and normalising by the mass creates the cumulative distribution function (CDF, often written as $p$) form of the NFW profile. The CDF form is well known in the literature, and available in a number of analytic forms. The inverse version of the CDF is the quantile distribution function (QDF, often written as $q$, where $q(p(x))=x$ for any distribution). By creating a fine grid of CDF samples as a function of concentration and radius it is possible to draw random Uniform samples and use the QDF to project these onto the desired radius distribution for the NFW. The {\sc Colossus} \citep{colossus} and {\sc astropy-halotools} \citep{astropy} packages use this technique\footnote{https://bdiemer.bitbucket.io/colossus/ and https://github.com/astropy/halotools/blob/master/halotools/ respectively}.\\
\end{enumerate}

For the second technique (projected Uniform sampling on the QDF), all published and known variants use numerical interpolation schemes or look-up tables to achieve the QDF inversion. In this short research note we detail the proof for a simple analytic scheme that allows for highly efficient and exact sampling of the NFW profile for any concentration.

\section{Derivation of the analytic CDF and QDF for the NFW profile}
We start with the NFW density profile at radius $q=R/R_{vir}$ (i.e.\ normalised to some fraction of the virial radius $R_{vir}$) for a halo of concentration $c$ (defined as $c=R_{vir}/R_{s}$, where $R_s$ is the standard definition of the scale radius):
\begin{equation}
\rho(q) \propto \frac{1}{qc(1+qc)^{2}}.
\end{equation}
The dark matter mass enclosed within radius $q$ is then given by
\begin{equation}
M(q) \propto \mathrm{ln}(1+qc)-\frac{cq}{1+cq}.
\label{eq:mass}
\end{equation}
When populating dark matter halos with satellite galaxies, a common assumption is that the galaxies follow the mass distribution of the halo. In this case, the cumulative probability, $p$ of finding a galaxy at radius $q$ is $p=M(q)/M(1)$. Following the standard method of drawing from a PDF, we seek to generate random values for $p\in [0,1]$ and invert the CDF. An analytic solution for the inversion can be obtained by substituting in Eq.~\ref{eq:mass}, performing the common mathematical trick of adding 1 to each side of the equation and taking the exponential:
\begin{align}
pM(1) + 1 &= \mathrm{ln}(1+cq) + \frac{1}{1+cq}, \\ 
e^{pM(1)+1} &= (1+cq)e^{\frac{1}{1+cq}}.
\label{eq:cdf1}
\end{align}
Although at first glance this equation seems impossible to solve for $q$, we can make use of the Lambert W function $W_{0}(y)$ which can be evaluated easily and efficiently by calling built-in routines in R, python (scipy) or using the GSL libraries in C. The best reference for details and applications of Lambert W is a research note provided by academics involved with the symbolic analysis software {\sc Maple}\footnote{https://cs.uwaterloo.ca/research/tr/1993/03/W.pdf}. Using this function allows us to solve equations of the form
\begin{equation}
y = xe^{\frac{1}{x}} \Rightarrow x=-[W_{0}(-y^{-1})]^{-1}.
\end{equation}
Comparing this to Eq~\ref{eq:cdf1}, and following some simplification we see that the solution is actually given by
\begin{equation}
q = -\frac{1}{c}\biggl[1+\frac{1}{W_{0}(-e^{-pM(1)-1})}\biggl].
\end{equation} 
Given the above, this opens up an analytic route for generating exact random samples of the NFW for any $c$:
\begin{equation}
r([0,1]; c) = q(p=U[0,1]; c).
\end{equation}
I.e.\ we can make random samples from 0--1 and use our solution for $q(p,c)$ to project this onto our normalised radius $q$, where in this context $q=R/R_{vir}$, but in practice it can be rescaled to be any radius the user desires. Care must be taken to re-interpret the correct meaning of $c$ if the outer radius is not being treated at $R_{vir}$, e.g.\ if you want to generate samples out the $2 R_{vir}$ with a classic $c=R_{vir}/R_s=5$ then you would generate samples using $c' = 2 c$ and multiply the returned $q$ from the random samples by $2 R_{vir}$.

\section{R and Python Tools for Random Sampling}

The authors have created two simple sets of tools for generating the samples using the above methodology in the popular languages {\sc R} and {\sc Python}. They are available from their maintained GitHub repositories immediately \footnote{{\sc R}: https://github.com/asgr/NFWdist; {\sc Python}: https://github.com/CullanHowlett/NFWdist}. The {\sc R} variant will be uploaded to CRAN in parallel to this paper. Given their use of standard functions and C-based GSL, it is trivial to port this methodology into any language, including {\sc C} and {\sc C++}.

\begin{figure}[htbp]
\begin{center}
\includegraphics[width=8cm]{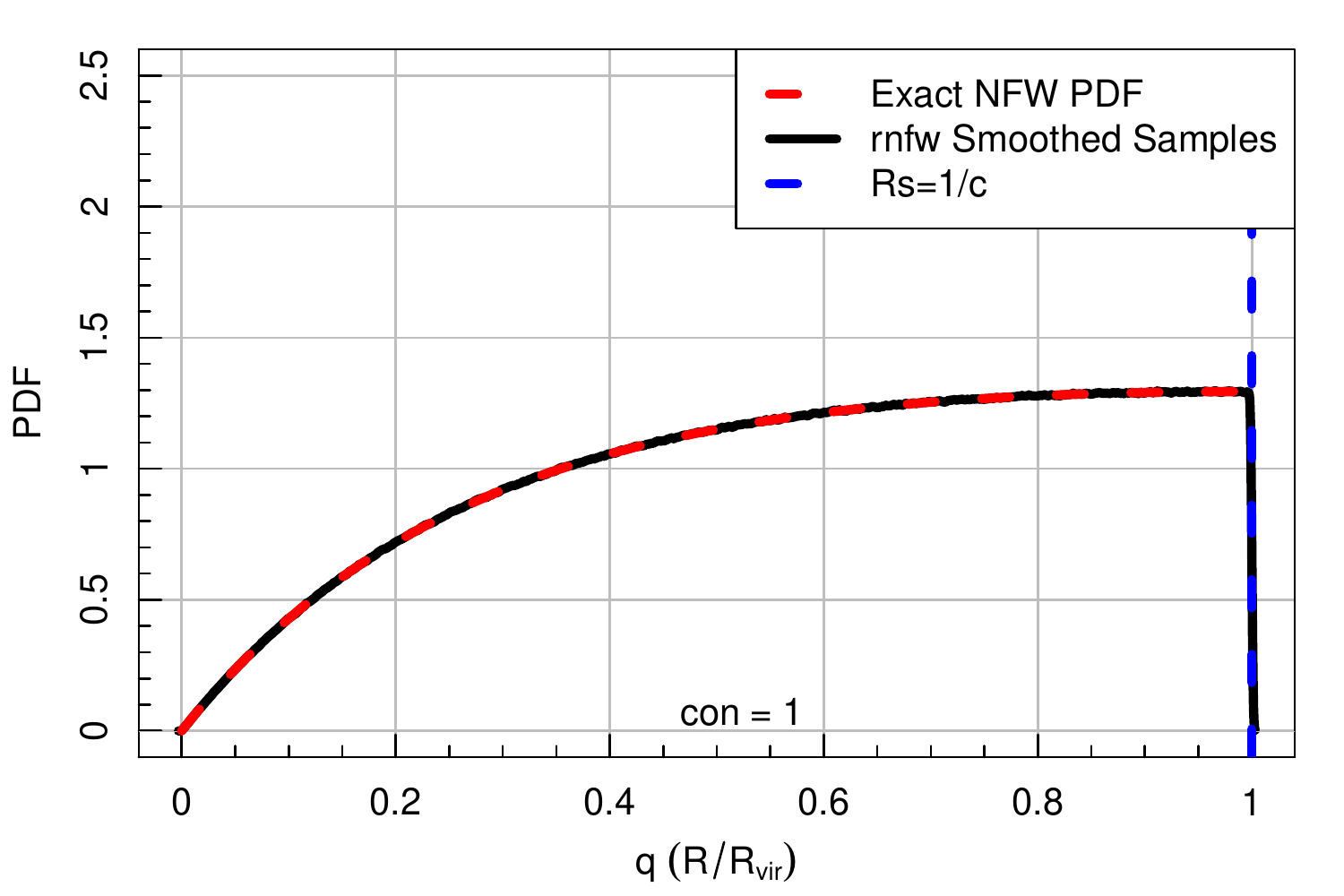} \includegraphics[width=8cm]{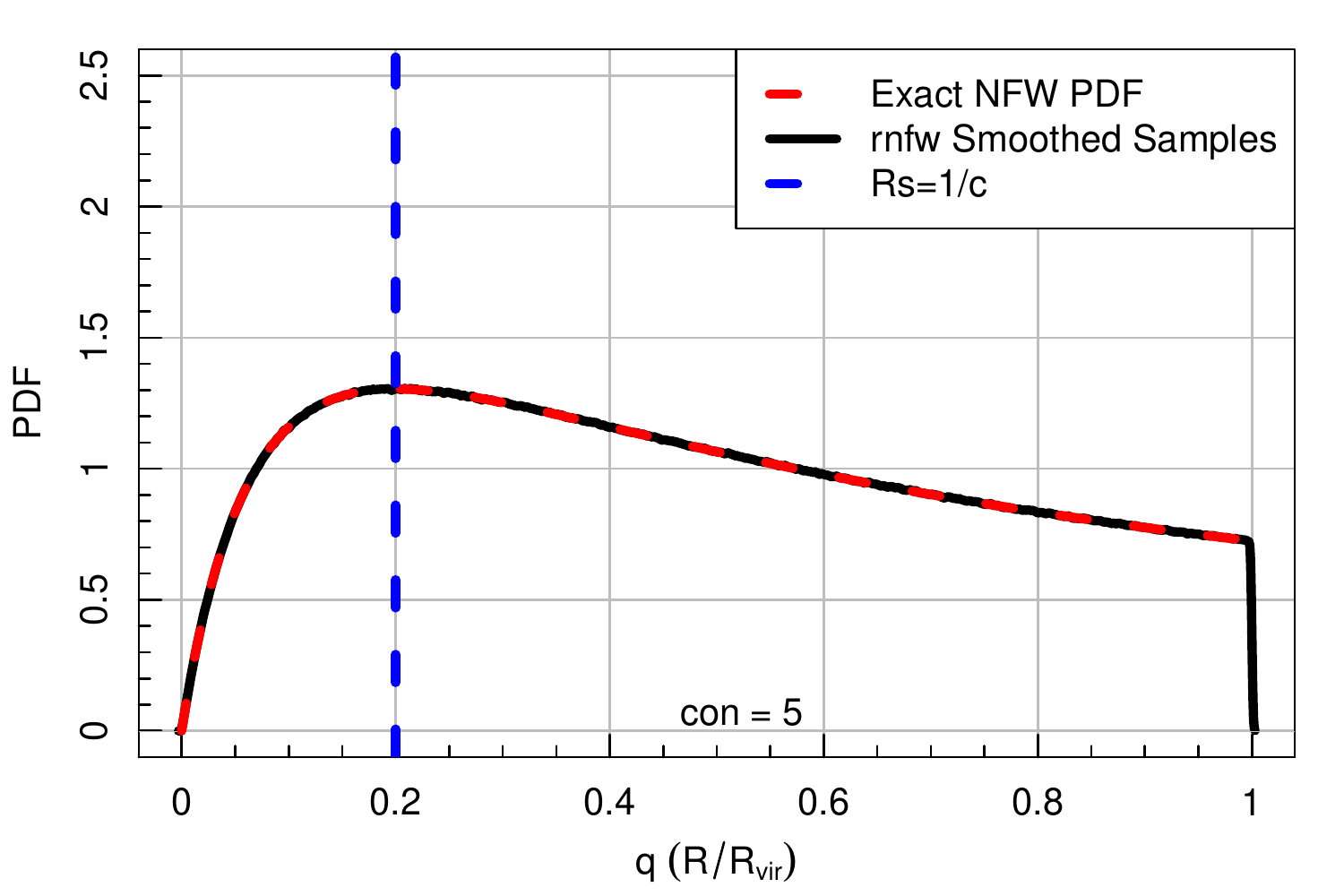}\\
\includegraphics[width=8cm]{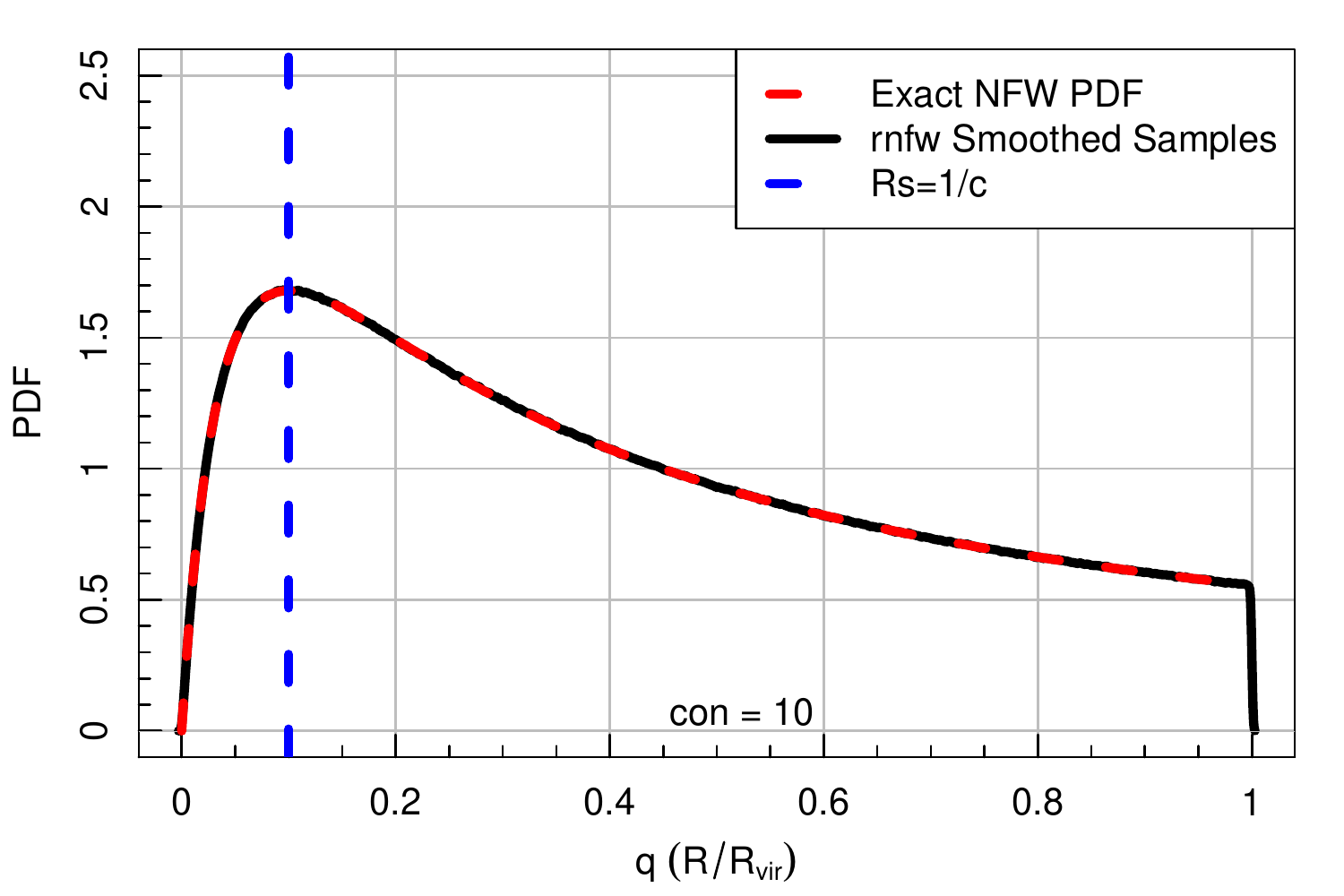} \includegraphics[width=8cm]{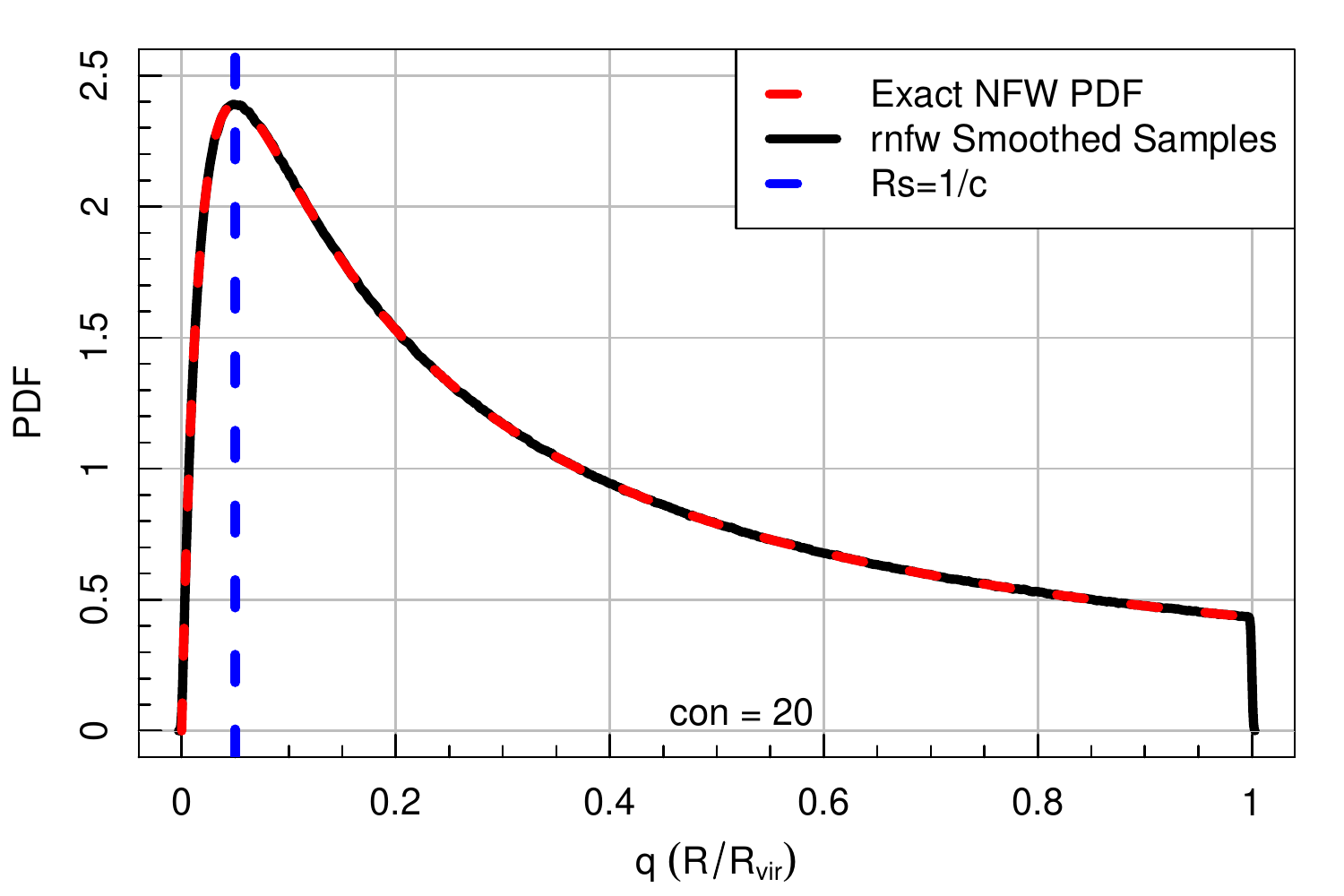}\\
\caption{Comparison of the exact PDF form of the 3D NFW profile and the smoothed (Normal bandwidth = 0.001) density profile of $10^8$ random samples for the same concentration (made using the {\sc R} {\tt rnfw} function included with the GitHub package provided with this note). The four plots show concentrations $c=1,5,10,20$. The small deviations seen at the distribution extremes are artefacts of the smoothing kernel Eddington bias which is particularly noticeable where there are discontinuities, i.e.\ at the sampling limits in this case. We should expect the PDF mode to appear at $R_s=1/c$, which they do in all cases.}
\label{fig:RNFW}
\end{center}
\end{figure}

Figure \ref{fig:RNFW} shows the outputs from the {\sc R} variant of the sampling code, where we compare the smoothed density of the samples to the analytic form of the PDF ($p$ above). They agree perfectly for all positive valued concentrations.

On a modern MacBook Pro with 4 cores, 10,000 samples takes $\sim$1 ms (for both {\sc R} and {\sc Python} implementations), which is approximately five times slower than the Uniform sampler that comes with base {\sc R} / {\sc Python}, and two times slower than the Normal sampler. Most of the time is spent computing the Lambert W function for the QDF, so faster implementations of this scheme would speed up any computation (the {\tt lamW} {\sc C++} implementation available for {\sc R} is the fastest the authors have come across to date, followed by the general GSL library available for {\sc C}). Sampling this way is roughly a factor four faster than reasonable implementations of the earlier accept-reject and numerical inversion schemes discussed (where the latter is harder to compare since the user must make a choice regarding accuracy). It is also exact.

\section{Acknowledgements}

Thanks to Claudia Lagos for the original conversation that sparked this work, and also to Danail Obreschkow, Rodrigo Tobar and Chris Power for useful conversations.

\end{document}